# Earth as a Proxy Exoplanet: Deconstructing and Reconstructing Spectrophotometric Light Curves

Lixiang Gu[1,2], Siteng Fan[2*], Jiazheng Li[2], Stuart Bartlett[2], Vijay Natraj[3], Jonathan H. Jiang[3], David Crisp[3], Yongyun Hu[1], Giovanna Tinetti[4], Yuk L. Yung[2,3]

## ABSTRACT

Point source spectrophotometric ("single-point") light curves of Earth-like planets contain a surprising amount of information about the spatial features of those worlds. Spatially resolving these light curves is important for assessing time-varying surface features and the existence of an atmosphere, which in turn is critical to life on Earth and significant for determining habitability on exoplanets. Given that Earth is the only celestial body confirmed to harbor life, treating it as a proxy exoplanet by analyzing time-resolved spectral images provides a benchmark in the search for habitable exoplanets. The Earth Polychromatic Imaging Camera (EPIC) on the Deep Space Climate Observatory (DSCOVR) provides such an opportunity, with observations of ~5000 full-disk sunlit Earth images each year at ten wavelengths with high temporal frequency. We disk-integrate these spectral images to create single-point light curves and decompose them into principal components (PCs). Using machine learning techniques to relate the PCs to six preselected spatial features, we find that the first and fourth PCs of the single-point light curves, contributing ~83.23% of the light curve variability, contain information about low and high clouds, respectively. Surface information relevant to the contrast between land and ocean reflectance is contained in the second PC, while individual land sub-types are not easily distinguishable (<0.1% total light curve variation). We build an Earth model by systematically altering the spatial features to derive causal relationships to the PCs. This model can serve as a baseline for analyzing Earth-like exoplanets and guide wavelength selection and sampling strategies for future observations.

[*]Corresponding author: stfan@gps.caltech.edu
[1]Department of Atmospheric and Oceanic Sciences, Peking University, Beijing, 100871, China;
[2]Division of Geological and Planetary Sciences, California Institute of Technology, Pasadena, CA 91125, USA;
[3]Jet Propulsion Laboratory, California Institute of Technology, Pasadena, CA 91109, USA;
[4]Department of Physics and Astronomy, University College London, Gower Street, WC1E 6BT London, UK.

# 1. INTRODUCTION

The search for life beyond Earth is one of the ultimate goals of planetary science, and significantly motivates the detection and characterization of exoplanets. Since the first exoplanet was detected (Campbell et al. 1988), 4000 more have been confirmed to date (NASA Exoplanet Archive; exoplanetarchive.ipac.caltech.edu). Several of these exoplanets are in habitable zones and have Earth-like properties (mass, radius, and composition), e.g., TRAPPIST-1 e, f, and g (Gillon et al. 2017). However, existing measurements and those in the near future are far from adequate to directly image an exoplanet with resolution higher than a single point. Analyzing single-pixel observables that can be obtained from a distance will remain the only approach for assessing the habitability of exoplanets in the foreseeable future. After decades of astrophysical and chemical characterization of exoplanets, astrobiological characterization is becoming increasingly important.

Planet Earth, the only celestial object known to harbor life, provides the single ground truth for a habitable world (Seager & Bains 2015). Analyzing the light curves of Earth, treated as a proxy exoplanet, can therefore provide a benchmark for evaluating the detectability and observation limit of potential signals from a habitable exoplanet. Cowan et al. (2009, 2011) and Cowan & Strait (2013) used two observations of multi-wavelength light curves of the Earth obtained by the Deep Impact spacecraft, each spanning one day, to analyze changes in surface features and clouds. They found that two characteristic spectra ("eigencolors"), which dominate the light curve variations, are sufficient to represent the spectral response of the Earth. However, due to the short time spans, these observations were not adequate for analyzing any long-term variations, which may have a large impact on the light curves.

The Earth Polychromatic Imaging Camera (EPIC) onboard the Deep Space Climate ObserVatoRy (DSCOVR) provides an opportunity to study long-term variations in time-resolved, multi-wavelength, photometric light curves of the sunlit disk of the Earth. By spatially integrating the disk images to simulate light curves of the Earth as a point source (hereafter called "single-point light curves"), Jiang et al. (2018) qualitatively analyzed the influence of different surface types and clouds on the light curves, and derived the minimal required sampling rate to retrieve the rotation period of the Earth. Fan et al. (2019) demonstrated that information about the surface is contained in the second principal component (PC) of the single-point light curves, and recovered a surface map in the presence of interference from clouds. Further improvements were made by Aizawa et al. (2020) by comparing different regularization methods. This technique can serve as a standard surface mapping strategy for future exoplanet studies (Fan & Yung 2020). However, each wavelength channel was given equal importance in these surface mapping efforts, resulting in a deviation from real observables. Also, given that correlation does not necessarily imply causation, validation is needed to confirm the proposed physical meaning of the PCs.

In this work, we perform a more sophisticated and in-depth analysis of the Earth as a proxy exoplanet by using a novel, multi-step methodology to reanalyze spectral images obtained by EPIC during the year 2016. First, we use the EPIC images to categorize pixels as one of six spatial features: ocean, desert, snow/ice, vegetation, low and high clouds. Second, we deconstruct the spectrophotometric light curves into their PCs using singular value decomposition (SVD), and quantitatively analyze the contribution of the six spatial features to the light curves and their temporal variation. Third, we use machine learning to interpret the physical meaning of the PCs and to relate them to the spatial features. Fourth, we develop an Earth model to reconstruct synthetic EPIC images by systematically altering spatial features, thereby obtaining a causal relationship between spatial features and PCs.

## 2. METHOD
### 2.1. DSCOVR/EPIC

The EPIC instrument on the DSCOVR satellite observes the Earth at a distance of ~ $1.5 \times 10^6$ km, from the first Lagrangian point (L1) of the Sun-Earth system. It images the full sunlit disk of the Earth using a 2048×2048 charge-coupled device (CCD) with ten narrow-band filters centered at 317, 325, 340, 388, 443, 551, 680, 688, 764, and 780 nm. The observations have a temporal resolution of ~68–110 minutes, which results in ~5000 images in each channel every year. Integrating these full-disk images provides single-point light curves of the Earth, which can then serve as a proxy exoplanet as seen at a geometry close to secondary eclipse. Combining the EPIC observations with known ground truth data, we can analyze the relationship between spatial features on the Earth and the corresponding light curves.

### 2.2. Earth Image Deconstruction

The light curve of an exoplanet is a time-dependent convolution of geometry and spatial feature information (Equation 1). Since exoplanets will be unresolved point sources for the foreseeable future, understanding the contribution of each spatial component to the light curve is critical for characterizing the habitability of exoplanets. EPIC level 2 products, which are derived using known properties of the Earth (e.g., surface type distribution, surface and atmospheric composition, spectra of spatial features), provide the ground truth of this work. Using spatially resolved EPIC-view land cover (DSCOVR EPIC L2 composite 01, doi:10.5067/EPIC/DSCOVR/L2_COMPOSITE_01) and cloud (DSCOVR EPIC L2 Cloud Products, doi:10.5067/EPIC/DSCOVR/L2_CLOUD_01) data, we derive the fractions of six different spatial features (ocean, desert, snow/ice, vegetation, low and high clouds) for every DSCOVR observation. Observations from multiple low Earth orbiting and geostationary satellites are combined to generate the EPIC-view composite data, including land cover fractions for twenty surface types for each pixel, which are then grouped into four major categories (Table 1).

The observed radiance can be written as a matrix $\mathbf{R_{[T*N]}}$, whose (t,n)-th element, $\mathbf{R_{tn}}$, is the disk-integrated reflected radiance for the t-th time step in the n-th channel. $\mathbf{R}$ is a weighted average of the reflected radiance from different spatial features:

$$\mathbf{r_{t[1*N]}} = \sum_{p=1}^{P} \mathbf{w_{tp[1*S]} X_{tp[S*N]}} \quad (1)$$

where $\mathbf{r_t}$ is the t-th row of $\mathbf{R}$, which refers to the observed reflection spectrum for the t-th time step; $\mathbf{w_{tp}}$ is a row vector, whose s-th element is the weight of the s-th spatial feature in the p-th pixel for the t-th time step; and $\mathbf{X_{tp}}$ is the reflection matrix, whose (s,n)-th element is the time- and geometry-dependent reflectivity in the n-th channel of the s-th spatial feature at the geometry of the p-th pixel for the t-th time step.

In this work, we classify clouds as low/high if they are located below/above 6 km, which is a typical separation between low and middle clouds in the Earth's atmosphere. We further separate the clouds by thickness since thin and thick clouds have different features in the spectra. Similar to the MODerate-resolution Imaging Spectroradiometer (MODIS) methodology (Platnick et al. 2003), the EPIC retrieval derives cloud optical thickness (COT) using the 680 nm channel. We assume low/high clouds to be water/ice clouds for simplicity. We further denote them as optically thin/thick clouds if the COT is below/above 10, in line with previous work (e.g., Robinson et al. 2011). Spectra of thin-cloud covered pixels are assumed to be linear combinations of spectra of thick clouds and those of the underlying surface. We find that weights of 36.3% and 63.7% for thick clouds and surface, respectively, result in the minimal mean square error (MSE) for the thin-cloud covered pixels.

In order to understand the influence of the different components, we deconstruct the light curves using SVD to obtain the PCs, or "eigencolors", of the Earth's spectrum. Unlike the methodology employed in Fan et al. (2019), the light curves are not normalized by the variance of each channel (Equation 2), since the amplitude of variation in each channel is an indicator of their relative spectral contributions. The SVD process can be expressed as follows:

$$R_{[T*N]} - \bar{R}_{[T*N]} = U_{[T*N]} \Sigma_{[N*N]} V^T_{[N*N]} \qquad (2)$$

where $\bar{R}$ is a matrix for the time-averaged spectrum, whose rows $\bar{r}_{t[1*N]}$ are equal to the averaged $r_t$ over the time index, t; $U$ is the time series matrix, whose n-th column is the time series of the n-th PC; $\Sigma$ is a diagonal matrix, whose (n,n)-th element is the n-th singular value from the decomposition; and $V$ is the eigenvector, or "eigencolor", matrix, whose n-th column is the spectrum corresponding to the n-th PC.

To ascribe physical meaning to the deconstructed spectrophotometric light curves, we follow the methodology of Fan et al. (2019), using a machine learning technique, the Gradient Boosted Regression Trees (GBRT; Friedman 2001), to relate each PC to spatial features. The training data for the GBRT model is the time series corresponding to the PCs from the decomposition (rows of U), while the training labels are the disk-integrated fractions of the spatial features, weighted by the cosine of solar zenith angle, which represents the incident solar flux. The model uses tree structures to classify the training data according to their values along the ten PC dimensions. A threshold is set up for the PCs each time to obtain the minimal MSE for the label prediction. Then, the relative importance of each PC can be computed as the number of times it serves as the threshold. This procedure is repeated for the six spatial features. Further details regarding the GBRT model can be found in Fan et al. (2019).

### 2.3. Earth Image Reconstruction

Instead of developing a forward model with sophisticated treatment of radiative transfer effects, we treat the reflected intensity of a given spatial feature at a given wavelength as a function of the solar and viewing zenith angles. We compute a two-dimensional moving average of the pixel reflectance in the phase space of these two angles. The phase angle is not considered as an independent variable since its variation over all observations is small (4-15 °). The reflected intensity of each spatial type is obtained directly from the categorized EPIC pixels. Therefore, all atmospheric effects (such as absorption and scattering) are contained in the geometry-dependent pixel reflectance; this can also be interpreted as the apparent brightness of a pixel observed at the DSCOVR location. Our Earth image model then uses a time-varying spatial feature map to reconstruct DSCOVR observations.

We first derive a benchmark of the Earth model by simulating DSCOVR observations with spatial feature distributions identical to the ground truth Earth, and generate corresponding single-point light curves; this is denoted as "synthetic data" hereafter. Then, we construct Earth models where some of the spatial features are altered; this allows emulation of Earth-like exoplanets with similar spectral features but different spatial feature coverage, and provides an opportunity for analyzing the causal relationship between PCs and spatial features. These simulations are designed to provide an insight into the physical meaning of different PCs. The six feature alteration experiments are summarized in Table 2. In these experiments, each spatial feature alteration is facilitated by replacements of one row in $X_{tp}$ by another in Equation 1. The experiments are the following: (a) No Low Clouds: all low clouds are removed, (b) Ocean World: all land is replaced by ocean, (c) No High Clouds: all high clouds are removed, (d) Flora World: all land is assumed to be vegetation, (e) Desert World: all land is assumed to be desert, and (f) No Clouds: both low and high clouds are removed. Here, we do not include an experiment for Snow/Ice World due to its similar spectral reflectance compared to that of clouds and its uncertain PC correlation, as described in Section 4. Through this process, seven (benchmark plus six altered) sets of simulated light curves are generated. The following analyses are based on these eight (observation plus seven simulated) datasets.

## 3. RESULTS

### 3.1. Light Curve Information Content

An example of reconstructed synthetic Earth images is shown in Figure 1a. It simulates the Earth with surface type distributions identical to the ground truth and real cloud coverage at 11:12 UTC, on August 15, 2016. All major spatial features of the Earth are recovered by the simulation. Disk-integration of these spectral images produces multi-wavelength synthetic light curves. A comparison between the observed light curves and the corresponding synthetic data for the year 2016 is shown in Figure 1b. The synthetic light curves agree well with observations at all wavelengths except for the first two ultraviolet (UV) channels, where the synthetic data shows smaller variation than the observations. This may be due to the seasonal variation of ozone in the atmosphere, which is averaged out in the spectral image reconstruction.

We then derive disk-averaged spectra of synthetic Earth assuming the entire sunlit disk is covered one at a time by each of the six spatial features (Figure 2a). Figure 2b shows the differences between these spectra and that of the average Earth, which is derived by averaging the synthetic data over the entire year. Some influence of the Earth's atmosphere can be identified in the spectral difference. The slope in the first three UV channels should originate from the combined effects of Rayleigh scattering and ozone absorption. Similarly, absorption due to the oxygen *A* and *B* bands at 764 nm and 688 nm, respectively, significantly affects the spectral shapes, resulting in a small spectral difference for all spatial features. The oxygen absorption feature is shallower for high clouds than it is for low clouds (or the surface) because high clouds reduce the optical pathlength for oxygen absorption more than low clouds. This indicates the feasibility of discriminating high and low cloud populations in disk-integrated spectra, as demonstrated in Section 3.2. Also, the spectral shape of ice/snow reflectance is very similar to that of low clouds; therefore, its contribution to the single-point light curves is difficult to be identified due to its much smaller coverage. Among the six spectra, vegetation is unique, showing monotonically increased reflectance at the four longest wavelengths, which is consistent with the vegetation red edge (VRE), and can in principle be detected.

We further investigate the contribution of each spatial feature to the light curves. Figure 2c shows the fractional reflectance contribution of these spatial features to the single-point light curves at the ten wavelengths. In the first four UV channels, the contributions from ocean and clouds (in addition to Rayleigh scattering and ozone absorption that are not considered in this analysis) are comparably dominant, with proportions of ~40% each. Atmospheric scattering and ozone absorption dominate the measured reflectance (Jiang et al. 2018), and land surface features are hard to disentangle from images at these wavelengths. At longer wavelengths, low clouds are always the dominant component of the reflection spectrum, with a contribution of up to ~60%. High clouds have relatively larger proportions at 688 and 764 nm (wavelengths with strong oxygen absorption) than the neighboring 680 and 780 nm channels (non-absorbing wavelengths), since the oxygen absorption has larger influence on the lower atmosphere and the surface due to higher column density above these spatial features. Vegetation shows higher contributions in the two near-infrared (NIR) channels, consistent with VRE. Seasonal cycles originating from the asymmetry between the northern and southern hemispheres are also noticeable in the time series, where more vegetation and less ocean appear in the boreal summer.

### 3.2. Light Curve Temporal Variation

Since exoplanets are unresolved point sources now and are likely to remain so in the foreseeable future, spatial information must be retrieved from the time-dependence of their spectrophotometric light curves. We use SVD to separate the contributions of different spatial features to the light curves. Figure 3a shows all ten eigenvalues derived from the SVD of the observed and synthetic data. In the construction of the synthetic light curves, only six spatial features are considered, and the variation of each spatial feature is averaged. This results in a loss of temporal variation and therefore smaller eigenvalues. The PCs can be classified into three groups according to the relative magnitude of the eigenvalues. The first two PCs (PC1 and PC2) contain more than ~97% of the total variance of the light curve temporal variations and have eigenvalues nearly one order of magnitude larger than the third. The next four PCs (PC3 to PC6) constitute the second group, which is expected to contain spatial information pertaining to minor spatial features above the noise level. We ignore the rest of the PCs since their contribution to the total variance is less than ~0.02%. Spectra of the first six PCs are shown in Figure 3b. Despite having slightly smaller eigenvalues, the synthetic single-point light curves successfully recover all the characteristic spectral features except for a switch of PC3 and PC4 between the observed and synthetic eigenvectors. This is a promising result given the significantly simplified reconstruction that consists of only six time-independent spatial features. The third and fourth eigenvalues of the observed light curves are very similar (~10% difference); it is easy to understand the PC switch given the approximate treatment of spatial features.

In order to interpret the meaning of each PC, we derive their correlations to the spatial features and evaluate their relative importance using the GBRT model (Figure 4). Linear correlations between the fraction of three spatial features and their corresponding PCs suggested by the GBRT model result are shown in Figure 5. Comparisons between the PC spectra and spectral differences corresponding to relevant spatial features (or combinations) from Figure 2b further illustrate the correlations (Figure 6). We then analyze the light curves from the six synthetic experiments with alteration of spatial features to validate the interpretation. Examples of a snapshot of Earth images in the 780 nm channel for the altered scenarios are shown in Figure 7.

PC1 contains information about low clouds. It has a flat spectral feature from the UV to the NIR (Figure 3b), which is consistent with the spectrum of clouds, and it shows a distinct dominant peak (relative importance of 70% for both observed and synthetic data) for the low clouds feature (Figure 4). Moreover, this large importance of PC1 does not occur for any of the other features. Linear correlation between the low cloud fraction and the PC1 time series is strong with a correlation coefficient of -0.8 (Figure 5a). The (scaled) spectral difference between low clouds and the averaged Earth (Figure 2b) also shows a clear high correlation with PC1 (Figure 6a). Above all, this is validated by the "No Low Clouds" simulations. The first eigenvalue decreases by a factor of 1.46 compared to the synthetic dataset when low clouds are removed (Figure 8a), and the original PC2 becomes dominant, which is indicated by the shape of the spectrum (Figure 9b), while the spectral features corresponding to the original PC1 disappear. Meanwhile, all eigenvalues other than the first increase, which may be due to the fact that the absence of low clouds enables clearer observation of surface features, enhancing the signal due to their temporal variation. Similar behavior can be seen in the "No Clouds" experiment (Figures 8f and 9b). In contrast, neither the first eigenvalue, nor the PC1 spectrum, has noticeable changes in any other experiments (Figures 8–10). This strongly supports the hypothesis that changes in low-cloud coverage have the dominant impact on PC1.

PC2 contains information about the major surface types: land and ocean. Its spectrum shows increasing reflectance with wavelength (Figure 3b), which is consistent with the spectral behavior of both land and ocean. The two surface types have exactly repeating periods, unlike clouds; hence, they are expected to appear in the same PC. The GBRT importance shows a peak in PC2 (relative importance >84%) when one of the surface types is set as the label, except for snow/ice (Figure 4). The exception for snow/ice may be due to its relatively low coverage on the Earth's surface (~10%), confinement to high latitudes (resulting in smaller weight for disk integration) and/or slow temporal change. The relationship between PC2 and land/ocean contrast is also confirmed with strong linear correlation. The ocean fraction and the PC2 time series are strongly correlated, with a correlation coefficient of -0.96 (Figure 5b). Comparison of PC2 with the spectral difference between land and ocean, where the land spectrum is computed as a weighted average of the vegetation and desert spectra, also shows a good agreement (Figure 6b). In the "Ocean World" experiment when all land is replaced by ocean, the second eigenvalue drops by almost one order of magnitude while the first stays the same (Figure 8b). Removal of land/ocean contrast results in the absence of surface variation, with the resultant loss of spectral features from the original data set in PC2 (Figure 10b). In the "Flora World" and "Desert World" experiments, which involve misrepresentation of land sub-types, the changes in the second eigenvalue are much smaller (<30%; Figures 8d and 8e), and the spectrum of PC2 remains similar to the original (Figure 10b). The eigenvalue changes in these two experiments indicate that the impact of individual land sub-types may show up in PC4, PC5 and PC6. Therefore, we conclude that PC2 contains information about the contrast in spectral features between land and ocean.

PC4 of the observed light curves (PC3 of the synthetic data) contains high cloud information. Since the third and fourth eigenvalues of the original light curves are close to each other (differences ~10%; Figure 3)), and the spatial variation is partially but differently averaged out during the reconstruction of the Earth images, these two PCs are switched between the original and synthetic eigenvectors, as suggested by both the PC spectra (Figure 3b) and the importance distribution (Figure 4). This PC presents a relatively flat spectrum except for a large spectral contrast in the two oxygen absorption channels (688 and 764 nm, Figure 3b) compared to neighboring non-absorbing channels, and shows a dominant importance (>71%) when the fraction of high clouds is set as the label in the GBRT model (Figure 4). The linear correlation between the high cloud fraction and the PC4 time series is strong with a coefficient of -0.82 (Figure 5c). PC4 spectrum has reasonable agreement with the spectral variation of high clouds, but it is not as good as those for PC1 and PC2 (Figure 6c). The reduced correlation is thought to be primarily due to the (arbitrary) cutoff between high and low clouds at 6 km, such that PC4 includes contributions from both altostratus clouds, which are composed primarily of liquid water droplets, and cirrus clouds, which are composed of ice particles. The different altitudes and scattering properties of these cloud types introduces subtle differences in their spectra. In the "No High Clouds" experiment, the switch of PC3 and PC4 between the experiment and synthetic spectra is evident (Figure 9c). The corresponding eigenvalue decreases by 51% (Figure 8c). All the other eigenvalues are largely unchanged from the original values (Figure 8c). Similarly, in the "No Cloud" experiment, the synthetic PC3 spectrum corresponded to the fifth eigenvalue, corroborating the decrease in significance of PC3 when high clouds are removed (Figure 9c).

The meaning of PC3 of the observed light curves is not clear based on our current analysis. It does not seem to be associated with any spatial features considered in this work since it does not show dominant relative importance for any of the labels (Figure 4). The steep slope of its spectrum in the UV (Figure 3b) suggests that it may be related to atmospheric Rayleigh scattering or ozone, which have dominant contributions in that spectral region. Proof or negation of this hypothesis needs an atmosphere radiative transfer model and/or high temporal frequency observations of ozone, both of which are beyond the scope of this work. PC5 shows a secondary peak when vegetation or desert fraction is set as the label, but not for any other cases (Figure 4); therefore, it may contain information about the contrast between vegetation and desert. The corresponding spectrum also shows an increase in reflectance with increasing wavelength in the NIR (Figure 3b), which is consistent with VRE. Moreover, in the feature alteration experiments where all land is assumed to be vegetation or desert, the fifth eigenvalue drops significantly (Figures 8d and 8e), and PC5 seems to be switched with PC6 (Figures 10e and 10f). This result presents a weak evident that vegetation information is contained in PC5.

## 4. DISCUSSION

Current state-of-the-art telescopes and those likely to be available in the foreseeable future cannot observe exoplanets at resolutions higher than one pixel. Analysis focusing on single-point photometric light curves, therefore, remains the only viable approach for characterizing spatial features on exoplanets. However, information about planetary surfaces and atmospheric features is encoded in time-resolved light curves. As long as the wavelengths of observation are sensitive to spectral contrasts between different spatial features, the surface and layers of clouds in the planetary system can be distinguished. In this analysis using the Earth as a proxy exoplanet, information about major surface types (land and ocean) is found in PC2, which agrees with the analysis in Fan et al. (2019) that used normalized light curves. Two layers of clouds are identified due to different oxygen column densities above the cloud tops, which is a new finding for exoplanet climate system characterization. The methodology used in this work holds even when generalized to light curve observations of real exoplanets that may have different surface and cloud compositions. Using SVD, different layers of spatial features can still be separated. As proposed in Fan et al. (2019), time series of PCs associated with surface features usually have regular envelopes and trends of variation, while those with clouds tend to be chaotic. These correlations between PCs and spatial features can be further confirmed with spectral analysis, since spectra of clouds are usually flat in the UV to the NIR, due to their large sizes compared to these wavelengths. This is shown as the PC1 and PC4 spectra for the case of the Earth (Figure 3b), and is equally valid for exoplanet clouds/haze.

Spectral features of gaseous species can not only enable detection of molecules on exoplanets; they can also be used for separating spatial features at different altitudes. The separation of two layers of clouds demonstrated in this work originates from the combined effect of oxygen absorption and cloud heights. Observations lacking gas absorption band measurements cannot resolve different layers of clouds. For illustration, we compare SVD and GBRT results with and without the two oxygen absorption bands (Figures 11 and 12). When the two oxygen bands are removed, the eigenvalues of the first three PCs show no noticeable difference (Figure 11a), except for an absence of the spectral dip due to oxygen absorption (Figure 11b). Meanwhile, PC4 is replaced by the original PC5, confirmed by both the value of eigenvalues and the shape of the spectra. (Figures 11a and 11b). This is a strong indication that the original PC4 arises from oxygen absorption. Further, the dominant importance of high clouds vanishes (Figure 12), implying that the spectra do not contain information about high clouds. These results clearly demonstrate that the two oxygen bands are critical for discrimination of different layers of clouds. On exoplanets, this can be any absorption band of any gas, as long as the gas column density is sufficiently different above the layers of clouds/haze.

In this work, we considered six spatial features, but are not able to clearly interpret PC3 of the Earth's light curve. It seems to have reasonable importance for high clouds, but its time series is not related to any labels we considered. Besides the possibility that it may be related to Rayleigh scattering or ozone, PC3 may also be a co-rotating eigenvector with PC4, and may hence share the same physical meaning. Our analysis shows that snow/ice are hard to recognize from the PCs due to their slow temporal change and preferred locations near poles, which may also be true for Earth-like exoplanets since changes in snow/ice cover usually follow seasonal cycles. Another possible reason is that snow and ice have similar spectral reflectance variations as clouds in the DSCOVR channels (Figure 2a); their signature may be overwhelmed by that of clouds. A recommendation for future exoplanet observations is to add more observations in the infrared to address this problem.

Some degeneracy exists in SVD. As SVD focuses on the deviation from mean values (Equation 2), the results only depend on spectral contrasts, which has no indication of the presence of any spatial features. In this work, for example, the existence of PC2 only reflect spectral contrast between land and ocean, but it does not indicate the co-presence of them. For future investigations of actual exoplanet observations, the bulk composition of the surface materials could, in principle, be inferred from the overall spectrum, while the compositions of the individual surface types can be inferred from deviations from the averaged spectrum. Another degeneracy of SVD is the arbitrary sign of time series of PCs. As SVD only evaluates projected variances (that are scalars) of the time series of light curves along the orthogonal axes, the positive directions of these axes are arbitrary. Therefore, it is possible that time series of PCs are negatively correlated with corresponding spatial features; this is, indeed, true for all three PCs analyzed in this work (Figure 5). Time series of PC1 and PC4 show negative correlations with low and high cloud fractions, respectively; the same is the case for the PC2 time series and ocean fraction.

Another spectral unmixing technique, non-negative matrix factorization (NMF), has also been applied to DSCOVR observations (Kawahara 2020). The NMF procedure also identified spatial features of land, ocean, and clouds from the multi-wavelength single-point light curves. A conceptual illustration of the difference between the SVD and NMF spectral decomposition approaches is provided in Figure 13. NMF is a non-orthogonal projection of the original light curves. The origin of the projected coordinate system stays the same, and the coefficients of all projected components are non-negative. In contrast, SVD is an orthogonal projection of the deviation from the mean. The origin of the new coordinate system moves to the mean of the original data points, and the signs of the eigenvector coefficients are arbitrary as discussed above. In the case of DSCOVR observations presented in this work, the standard deviation of the light curve in each channel is less than 10% of the mean value (Fan et al. 2019). In other words, the observation data points cluster in a small part of the original phase space. Therefore, the resulting NMF components, i.e., the directions of the projected coordinate axes, tend to be similar to each other, as shown in Kawahara (2020). The SVD technique performs better in terms of separating data points in this particular case. In other cases, NMF may be preferred due to its advantage derived from decomposing the original light curves, whose magnitude is usually larger than the deviation from the mean, which may ameliorate effects due to clustering of data points.

## 5. CONCLUSIONS

We study the Earth as a proxy exoplanet using observations of full-disk images made by DSCOVR/EPIC. Spatial-temporal variations of the multi-wavelength single-point light curves are analyzed. We deconstruct and reconstruct the DSCOVR spectral images using six spatial features. SVD is used to separate the contributions of the different spatial features. We use the GBRT model to derive correlations between the ground truth spatial features and the PCs, and evaluate possible causal relationships underlying these correlations by generating synthetic light curves with modified spatial features. The importance distributions generated by GBRT suggest that PC1, PC2 and PC4 contain information about low clouds, land/ocean spectral reflectance contrast and high clouds, respectively. PC3 may be associated with atmospheric Rayleigh scattering or ozone absorption, and PC5 is suspected to be indicative of vegetation (or lack thereof). The six synthetic experiments of an altered Earth demonstrate the validity of these interpretations. The results also show that SVD can obtain the original and altered surface information in spite of the presence of clouds, which is critical in the context of exoplanet observation. Moreover, two layers of clouds are found to be distinguishable through the bands of atmospheric absorption. This indicates that, even when the exoplanet cloud and surface composition is different from the Earth, spatial features of the surface and multiple layers of clouds can in principle be derived from its time-varying light curves.

## ACKNOWLEDGEMENTS

LG and YH are supported by the National Natural Science Foundation under grants 41530423, 41888101, and 41761144072. LG is also partly supported by the China Scholarship Council. A portion of this research was carried out at the Jet Propulsion Laboratory, California Institute of Technology, under a contract with the National Aeronautics and Space Administration (80NM0018D0004). DC and YLY acknowledge support from the Virtual Planetary Laboratory at the University of Washington. YH is also supported by the research project of Technology of Space Telescope Detecting Exoplanets and Life (D030201). GT was supported by the Science and Technology Funding Council of UK. JHJ, SB and YLY acknowledge funding support from the NASA Exoplanet Research Program NNH18ZDA001N.

TABLES

**Table 1.** Surface type categorization.

| | Index | Surface Type | | Index | Surface Type |
|---|---|---|---|---|---|
| **Vegetation** | 1 | Evergreen Needleleaf Forest | **Vegetation** | 11 | Permanent Wetlands |
| | 2 | Evergreen Broadleaf Forest | | 12 | Croplands |
| | 3 | Deciduous Needleleaf Forest | | 14 | Cropland Mosaics |
| | 4 | Deciduous Broadleaf Forest | | 18 | Tundra |
| | 5 | Mixed Forest | **Desert** | 13 | Urban and Built-up |
| | 6 | Closed Shrublands | | 16 | Bare Soil and Rocks |
| | 7 | Open Shrublands | **Ocean** | 17 | Water Bodies |
| | 8 | Woody Savannas | **Snow/Ice** | 15 | Snow and Ice (permanent) |
| | 9 | Savannas | | 19 | Snow (seasonal over land) |
| | 10 | Grasslands | | 20 | Ice (on water bodies) |

**Note.** The twenty surface types in the DSCOVR composite data are regrouped into four.

**Table 2.** Summary of the eight datasets.

|  | Spatial Features | | | | | |
| --- | --- | --- | --- | --- | --- | --- |
|  | **Ocean** | **Vegetation** | **Desert** | **Snow/Ice** | **High Clouds** | **Low Clouds** |
| **Observation** | O | O | O | O | O | O |
| **Synthetic** | S | S | S | S | S | S |
| **No Low Clouds** | S | S | S | S | S | Surface features below (S) |
| **Ocean World** | S | Ocean (S) | Ocean (S) | Ocean (S) | S | S |
| **No High Clouds** | S | S | S | S | Surface features below (S) | S |
| **Flora World** | S | S | Vegetation (S) | Vegetation (S) | S | S |
| **Desert World** | S | Desert (S) | S | Desert (S) | S | S |
| **No Clouds** | S | S | S | S | Surface features below (S) | Surface features below (S) |

**Note.** Spatial features are shown as different columns, and datasets as rows. Observed and synthetic data are denoted by "O" and "S", respectively. For the six alteration experiments, spatial features replacing the original synthetic data are indicated.

FIGURES

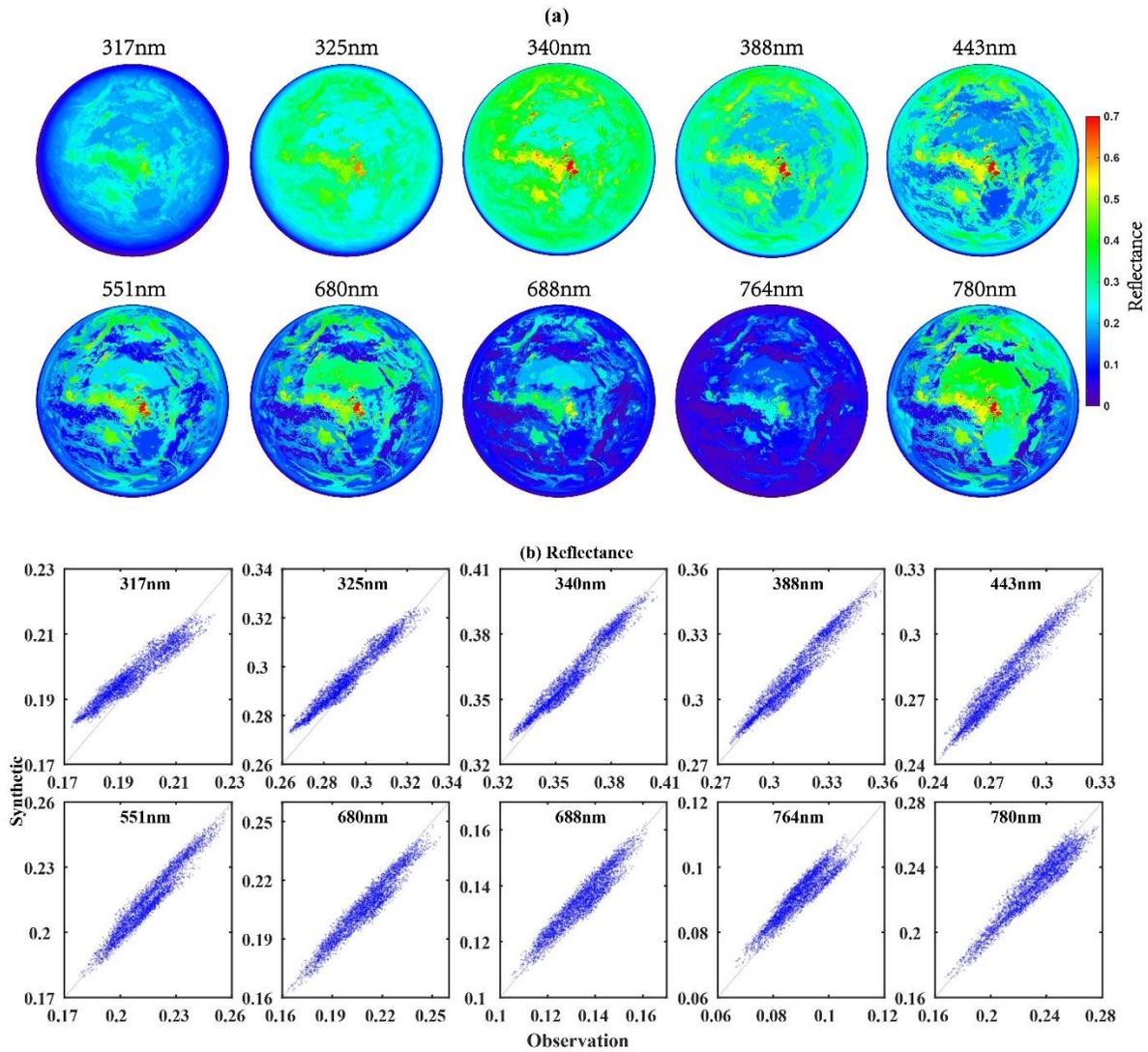

**Figure 1.** (a) Synthetic reflectance images in the ten DSCOVR channels at 11:12 UTC, August 15, 2016. (b) Disk-integrated reflectance derived from synthetic and observed data for the year 2016.

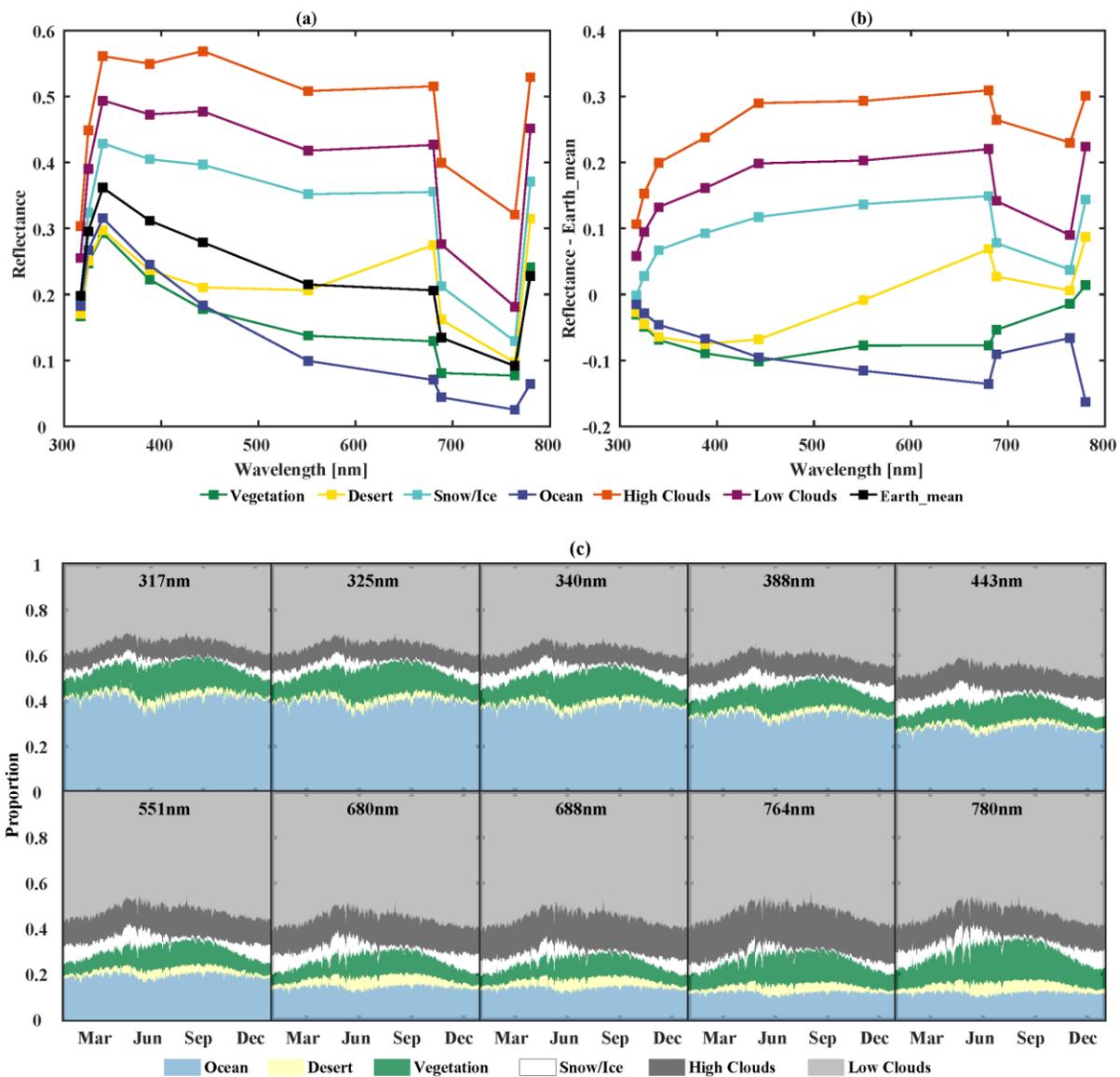

**Figure 2.** (a) Disk integrated full-phase reflection spectra for six spatial features (colored solid lines with squares) and that for the averaged Earth (black solid line with squares). (b) Difference between the reflection spectra for the spatial features shown in (a) and that for the averaged Earth. (c) Time series of daily averaged fractional reflectance contributions of the six spatial features.

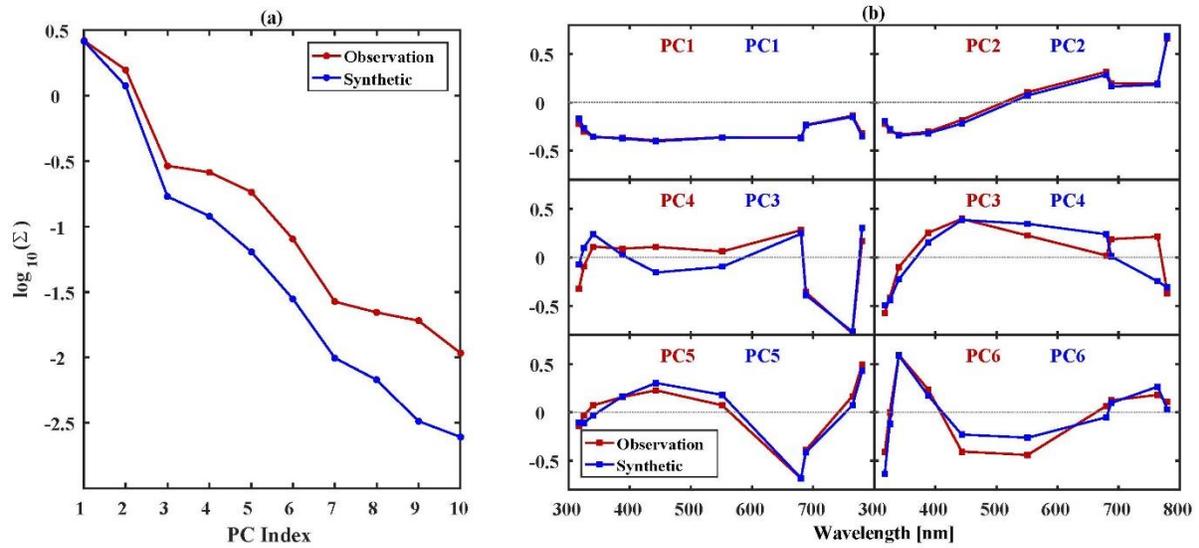

**Figure 3.** (a) Eigenvalues of observed (red) and synthetic (blue) light curves. (b) Eigenvectors (spectra) of the first six PCs for observed (red) and synthetic (blue) light curves. The text in each panel represents the PC index of the two datasets. PC3 and PC4 of observed data are switched to match the synthetic eigenvectors.

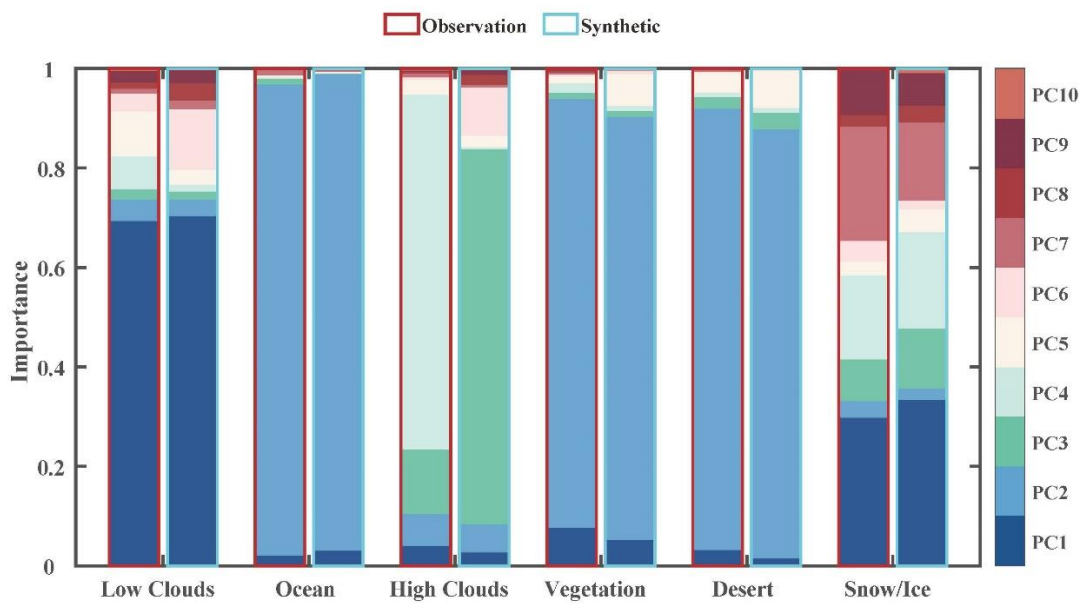

**Figure 4.** Relative importance of PCs (color-filled stacked bars) to explain each spatial feature. Bars with red edge denote results for observed data, while those with cyan edge are for synthetic data. The different fill colors indicate different PCs.

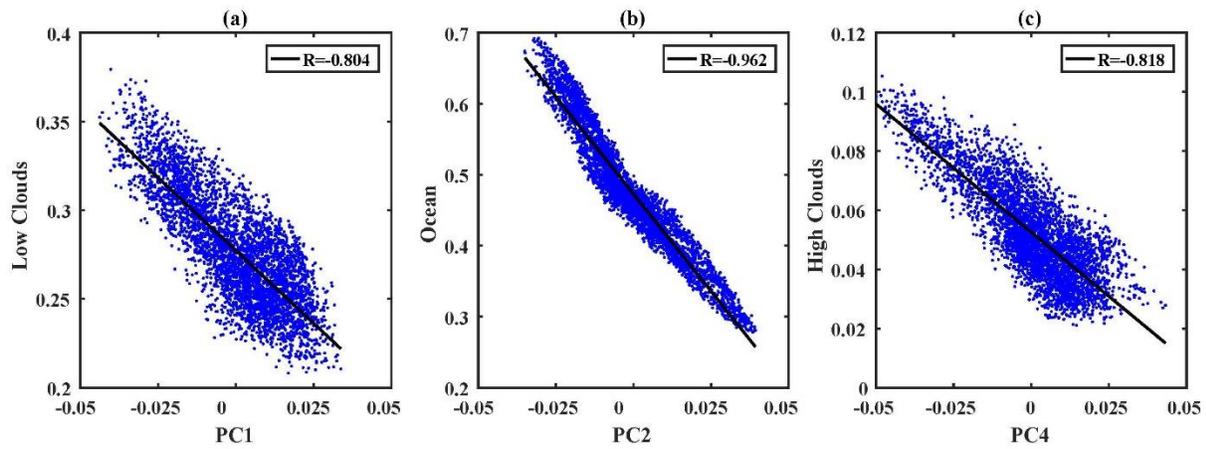

**Figure 5.** Scatter plots of PC time series against spatial feature fractions (blue dots): (a) PC1 and low clouds, (b) PC2 and ocean, and (c) PC4 and high clouds. The best linear-fit solutions are shown by black lines, with the correlation coefficients in the legend.

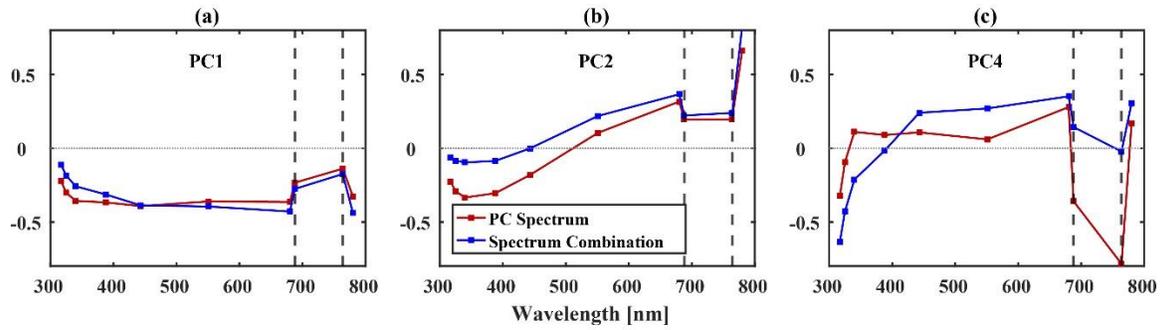

**Figure 6.** Comparison between eigenvectors of (a) PC1, (b) PC2, and (c) PC4 (red), and spectral differences corresponding to relevant spatial features (or combinations) from Figure 2b (blue). The two oxygen bands are shown as black dashed lines.

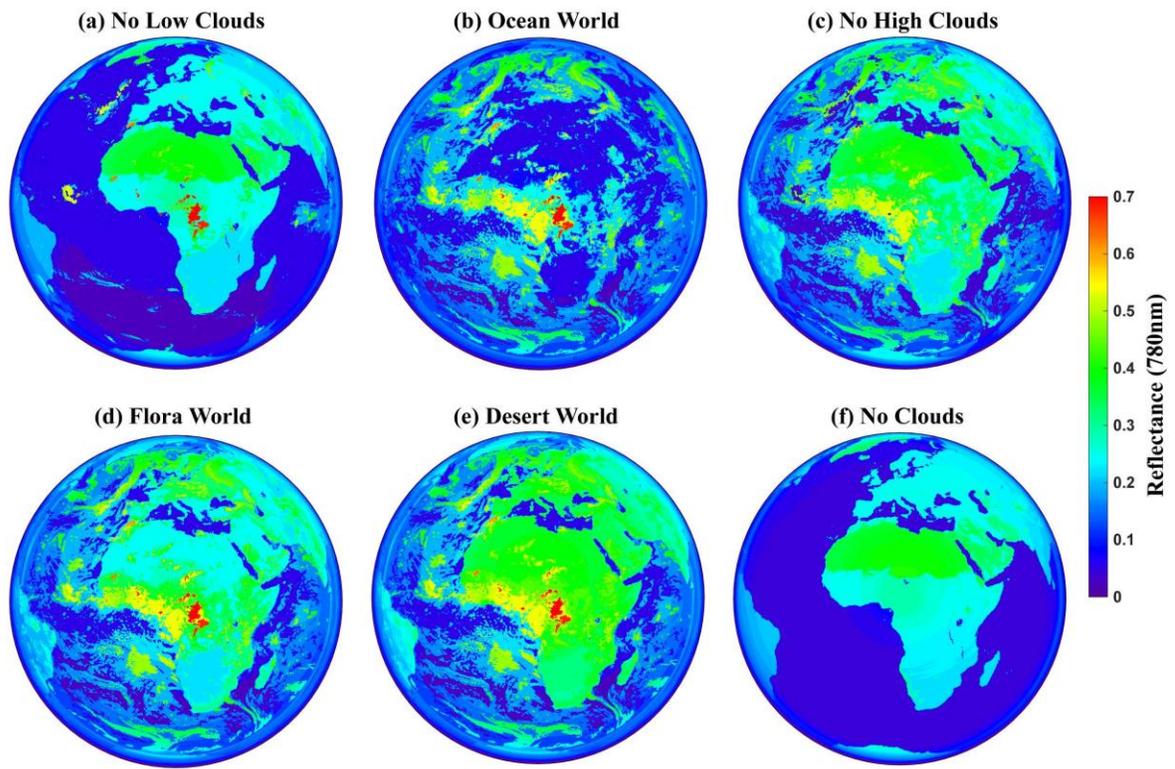

**Figure 7.** Synthetic reflectance images in the 780 nm channel at 11:12 UTC, August 15, 2016 for feature alteration experiments: (a) No Low Clouds, (b) Ocean World, (c) No High Clouds, (d) Flora World, (e) Desert World, (f) No Clouds.

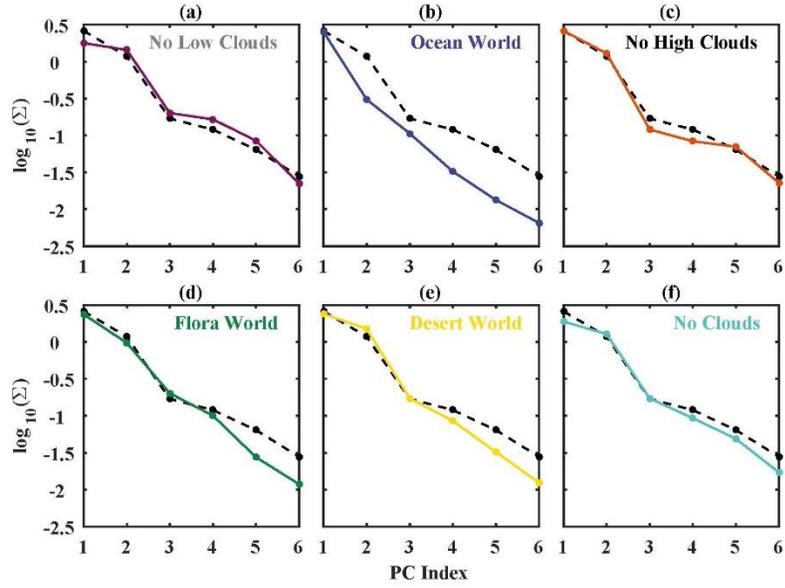

**Figure 8.** First six eigenvalues from feature alteration experiments: (a) No Low Clouds, (b) Ocean World, (c) No High Clouds, (d) Flora World, (e) Desert World, (f) No Clouds. Black dashed lines denote synthetic data (same as blue solid lines in Figure 3a). Solid lines denote data with altered spatial features.

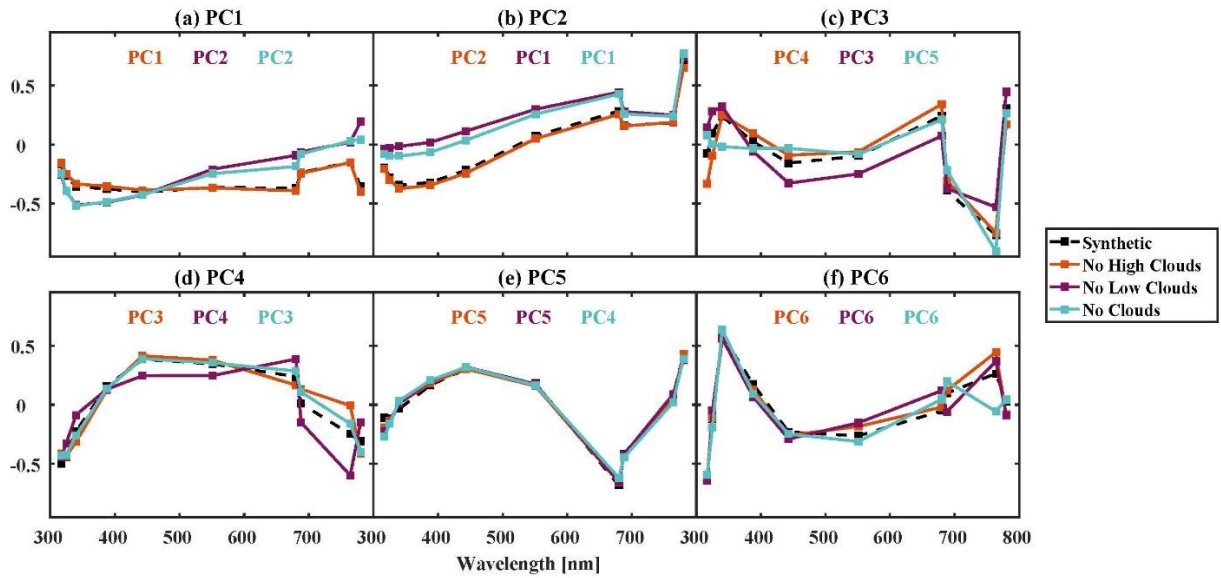

**Figure 9.** Eigenvectors (spectra) of (a) PC1, (b) PC2, (c) PC3, (d) PC4, (e) PC5, (f) PC6 from three feature alteration experiments: No High Clouds (orange), No Low Clouds (plum), and No Clouds (teal). Black dashed lines denote results for synthetic data. The PC indices for the different experiments are denoted by the text in each panel. For the No High Clouds case, PC3 and PC4 are switched. For the No Low Clouds case, PC1 and PC2 are switched. For the No Clouds case, the order of the first five PCs in (a)–(e) is PC2, PC1, PC5, PC3, and PC4.

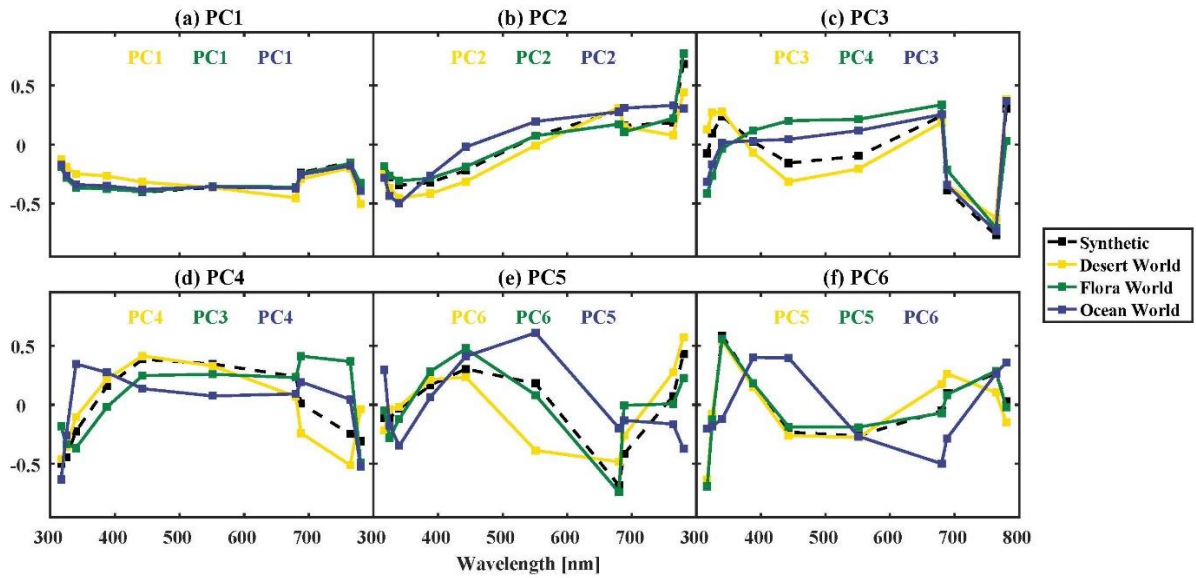

**Figure 10.** Same as Figure 9, but for the other three feature alteration experiments: Desert World (yellow), Flora World (green), and Ocean World (purple). For the Desert World case, PC5 and PC6 are switched. For the Flora World case, PC3 and PC4 are switched, as are PC5 and PC6.

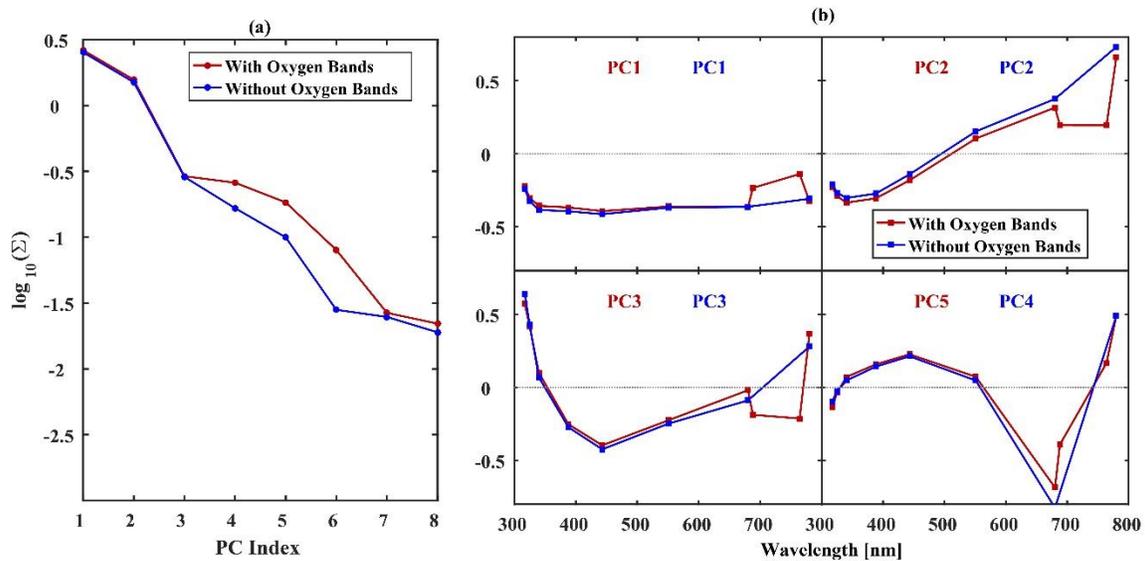

**Figure 11.** Same as Figure 3, but for observations with (red) and without (blue) the two oxygen bands. Note that PC5 for observations with oxygen bands match with PC4 for observations without those bands.

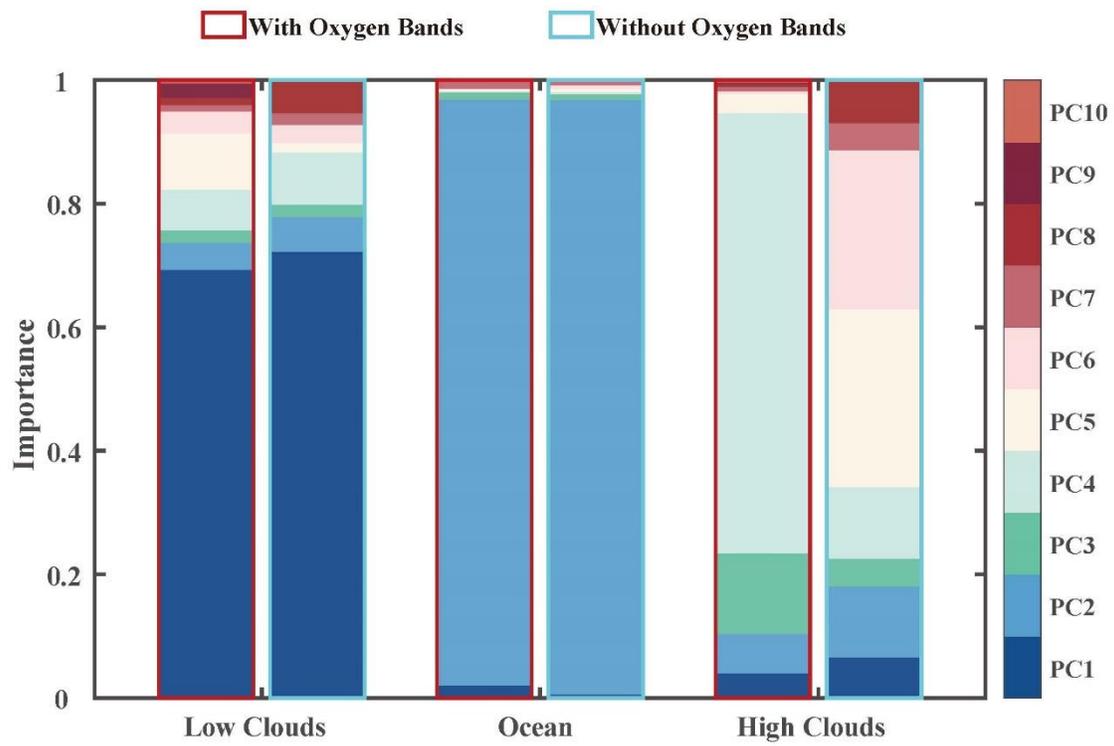

**Figure 12.** Same as Figure 4, but for observations with (bars with red edge) and without (bars with cyan edge) the two oxygen bands.

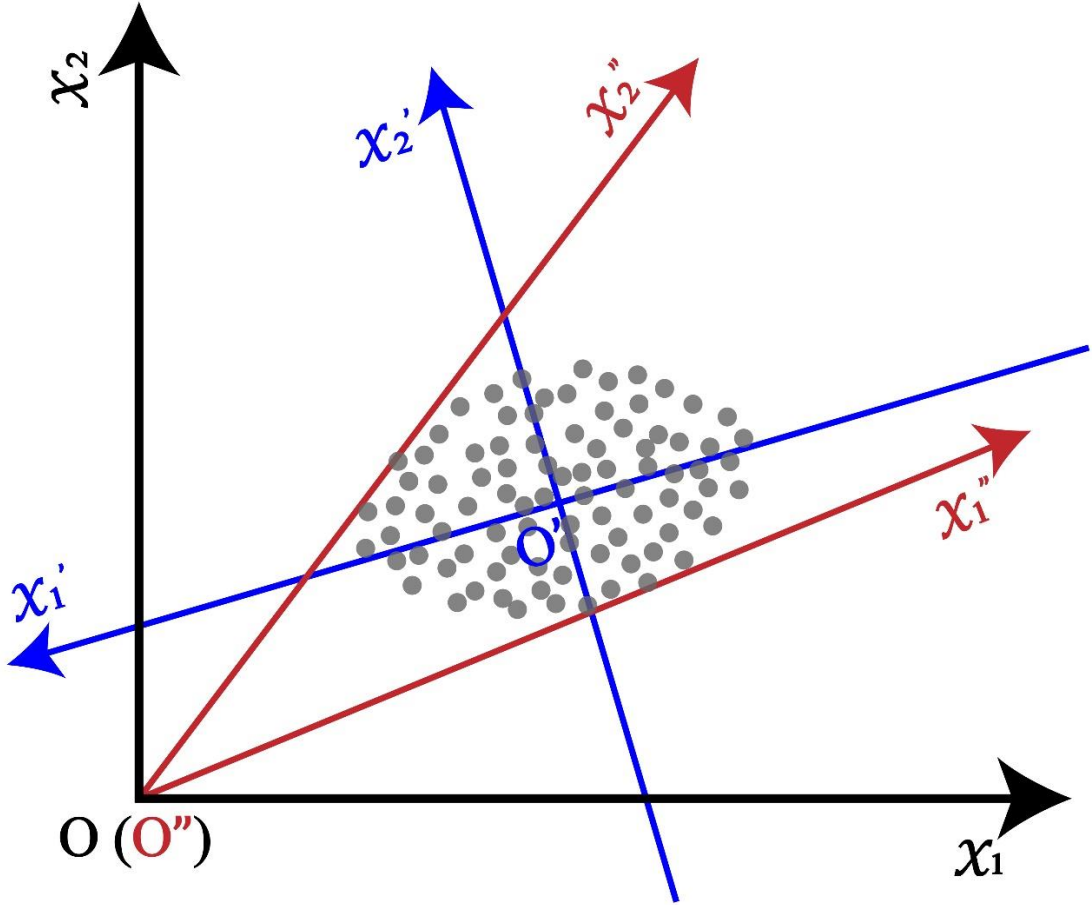

**Figure 13.** Conceptual illustration of the singular value decomposition (SVD) and non-negative matrix factorization (NMF) spectral unmixing techniques. The black coordinate system with origin O denotes the original phase space spanned by the data points (grey dots), while the blue and red coordinate systems denote projections obtained by SVD (origin O') and NMF (origin O''), respectively.